\documentclass[pre,twocolumn,showpacs,preprintnumbers,amsmath,amssymb]{revtex4}

\usepackage{graphicx}
\usepackage{dcolumn}
\usepackage{bm,amssymb}

\begin{document}

\newcommand{\bea}{\begin{eqnarray}}
\newcommand{\eea}{  \end{eqnarray}}
\newcommand{\bit}{\begin{itemize}}
\newcommand{\eit}{  \end{itemize}}

\newcommand{\be}{\begin{equation}}
\newcommand{\ee}{\end{equation}}
\newcommand{\ra}{\rangle}
\newcommand{\la}{\langle}
\newcommand{\U}{\widetilde{U}}


\def\bra#1{{\langle#1|}}
\def\ket#1{{|#1\rangle}}
\def\bracket#1#2{{\langle#1|#2\rangle}}
\def\inner#1#2{{\langle#1|#2\rangle}}
\def\expect#1{{\langle#1\rangle}}
\def\e{{\rm e}}
\def\proj{{\hat{\cal P}}}
\def\tr{{\rm Tr}}
\def\H{{\hat H}}
\def\Hdag{{\hat H}^\dagger}
\def\Lop{{\cal L}}
\def\Ehat{{\hat E}}
\def\Edag{{\hat E}^\dagger}
\def\Shat{\hat{S}}
\def\Sdag{{\hat S}^\dagger}
\def\Ahat{{\hat A}}
\def\Adag{{\hat A}^\dagger}
\def\U{{\hat U}}
\def\Udag{{\hat U}^\dagger}
\def\Zhat{{\hat Z}}
\def\Phat{{\hat P}}
\def\Op{{\hat O}}
\def\id{{\hat I}}
\def\x{{\hat x}}
\def\P{{\hat P}}
\def\Px{\proj_x}
\def\Pr{\proj_{R}}
\def\Pl{\proj_{L}}


\title{Short periodic orbit approach to resonances and the fractal Weyl law}


\author{J M Pedrosa$^1$, D Wisniacki$^2$, G G Carlo$^1$ and M Novaes$^4$}
\address{$^1$Departamento de F\'\i sica, CNEA, Av. Libertador 8250, Buenos Aires C1429BNP, Argentina\\ $^2$Departamento de F\'\i sica, 
FCEyN, UBA, Ciudad Universitaria, Buenos Aires C1428EGA Argentina\\$^4$Departamento de F\'\i sica, Universidade Federal de S\~ao Carlos, 
S\~ao Carlos, SP, 13565-905, Brazil}

\date{\today}

\begin{abstract}

We investigate the properties of the semiclassical short periodic orbit approach for the
study of open quantum maps that was recently introduced in [M. Novaes, J.M. Pedrosa, D.
Wisniacki, G.G. Carlo, and J.P. Keating, Phys. Rev. E {\bf 80}, 035202(R) 2009]. We
provide conclusive numerical evidence, for the paradigmatic systems of the open baker and
cat maps, that by using this approach the dimensionality of the eigenvalue problem is
reduced according to the fractal Weyl law. The method also reproduces the projectors
$|\psi^R_n\rangle\langle\psi_n^L|$, which involves the right and left states associated
with a given eigenvalue and is supported on the classical phase space repeller.

\end{abstract}

\pacs{05.45.Pq}

\maketitle

\section{Introduction}
\label{sec1}

One of the cornerstones of the semiclassical approach to quantum mechanics is the
Gutzwiller trace formula, which relates the fluctuating part of the quantum density of
states to the classical periodic orbits of a chaotic system \cite{Houches}. However, this
relation requires infinitely long orbits and involves divergent sums. More recently,
periodic orbits have been used to study quantum spectra in a different way, by
constructing some special quantum states, called scar functions, which are adapted to the
system's classical dynamics and provide a suitable basis for diagonalizing the
Hamiltonian. This formulation has the advantage of using only a small number of short
periodic orbits, and shedding light on the phenomenon of scarring \cite{Heller}, which is
an anomalous localization of stationary wave functions along periodic orbits.

In quantum scattering a prominent role is played by resonances or quasibound states.
These are eigenfunctions of the system with complex energy, whose (negative) imaginary
part is interpreted as a decay rate. In chaotic systems the number of states with a
prescribed decay rate grows as a power of the energy which is conjectured to be related
to a fractal dimension of the classical repeller, the set of initial conditions which
remains trapped in the scattering region for all, positive and negative, times. This
fractal Weyl law has been investigated in several systems
\cite{LuPrl03,SchomerusPrl04,Non1,Shepelyansky08,pedrosa1,Boro1,Hard3D}. One the other
hand, the (right) eigenfunctions are supported by the unstable manifold of this repeller \cite{ShepelyanskyPhD99,KeatingPrl06},
and its scarring properties have also been under investigation
\cite{Wiersig,carlo,Microlasers}.

The scattering analog of the short periodic orbit approach has been introduced in
\cite{p1}. Scar functions were constructed which are concentrated on the periodic orbits
but also extend along the unstable manifolds by means of a dynamical evolution up to the
order of the system's Ehrenfest time. This basis is adequate for the calculation of the
small fraction of resonances which have small decay rates, usually the most important
ones, without having to consider the multitude of rapidly decaying states. Also, it
provides a tool for studying localization on classical structures more directly, and
allows a natural semiclassical approach to resonance wave functions.

This method was used in \cite{p1} to reproduce the main resonances of an open baker map
for a specific value of $\hbar$. In the present work we focus on the scaling of the
method and show that it is compatible with the fractal Weyl law, not only for the baker
but also for a cat map, which is more generic. We also discuss the ability of method to
reproduce eigenfunctions. We find that a mixed quantity involving both types of
functions (right and left), which has been introduced in \cite{ermannPrl09}, can be accurately reproduced. The paper is organized as
follows. Section \ref{sec2} is focused on reviewing the quantum and the classical
versions of the systems that we have used in our study: the open baker and cat maps. In
Section \ref{sec3} we discuss the method and its results. Finally, we present our
conclusions in Section \ref{sec5}.

\section{Open quantum maps}
\label{sec2} Maps are paradigmatic systems in classical and quantum chaos because of
their simplicity \cite{Ozorio 1994,Hannay 1980,Espositi 2005}. We will consider
maps defined on the torus. When quantizing them, boundary conditions must be imposed for
both the position and momentum representations. This amounts to taking
$\bracket{q+1}{\psi}\:=\:e^{i 2 \pi \chi_q}\bracket{q}{\psi}$, and
$\bracket{p+1}{\psi}\:=\:e^{i 2 \pi \chi_p}\bracket{p}{\psi}$, with $\chi_q$, $\chi_p \in
[0,1)$. This implies a Hilbert space of finite dimension $N=(2 \pi \hbar)^{-1}$, and the
semiclassical limit is approached for large $N$. The system's propagator becomes a
$N\times N$ matrix. The discrete set of position and momentum eigenstates is given by
$\ket{q_j}\:=\:\ket{(j+\chi_q)/N}$ and $\ket{p_j}\:=\:\ket{(j+\chi_p)/N}$ with
$j\in\{0,\ldots, N-1\}$. They are related by a discrete Fourier transform, i.e.
$\bracket{p_k}{q_j}\:=\: \frac{1}{\sqrt{N}} e^{-2i\pi(j+\chi_q)(k+\chi_p)/N} \: \equiv \:
(G^{\chi_q, \chi_p}_N)$.

We consider a certain region of the torus to be the opening, through which particles can
escape. Its quantization is implemented by means of a projection operator $P$ on its
complement. We always choose a finite strip parallel to the $p$ axis and corresponding to
a range of $q$ values, so that the projector $P$ is quite simple in position
representation. If $U$ is the propagator for the closed system, then $\widetilde{U}=PUP$
represents the open one. It will have $N$ right eigenvectors $|\Psi^R_n\ra$ and $N$ left
ones $\la \Psi_n^L|$, which are orthogonal among themselves $\la
\Psi_n^L|\Psi^R_m\ra=\delta_{nm}$. Their norm is arbitrary, but one may choose $\la
\Psi_n^R|\Psi^R_n\ra=\la \Psi_n^L|\Psi^L_n\ra$.

The classical (tri)baker map
\begin{equation}
\mathcal B(q,p)=\left\{
  \begin{array}{lc}
  (3q,p/3) & \mbox{if } 0\leq q<1/3 \\
  (3q-1,(p+1)/3) & \mbox{if } 1/3\leq q<2/3\\
  (3q-2,(p+2)/3) & \mbox{if } 2/3\leq q<1\\
  \end{array}\right.
\label{classicaltribaker}
\end{equation}
is an area-preserving, uniformly hyperbolic, piecewise-linear and invertible map with
Lyapunov exponent $\lambda=\ln{3}$. Following \cite{Saraceno1,Saraceno2}, the quantum
version is defined in terms of the discrete Fourier transform in position representation
as
\begin{equation}\label{quantumbaker}
 U^{\mathcal{B}}=G_{N}^{-1} \left(\begin{array}{ccc}
  G_{N/3} & 0 & 0\\
  0 & G_{N/3} & 0\\
  0 & 0 & G_{N/3}\\
  \end{array} \right),
\end{equation}
where antiperiodic boundary conditions are imposed, $\chi_q=\chi_p=1/2$. For this system
we always take the opening as the region $1/3< q<2/3$.

The classical cat maps are of the form
\begin{equation}
\left(\begin{array}{c}q' \\p'\end{array}\right)=\mathcal{C} \left(\begin{array}{c}q \\p\end{array}\right){\rm mod \,} 1=
\left(\begin{array}{cc}
          c_{11}   & c_{12} \\
          c_{21}   & c_{22} \\
\end{array}\right)\left(\begin{array}{c}q \\p\end{array}\right){\rm mod \,} 1,\label{cat}
\end{equation}
where the $c_{ij}$ must be integers to ensure continuity and the conditions
Tr$\mathcal{C}>2$ and $\det\mathcal{C}=1$ are imposed to make the map hyperbolic and
area-preserving. Here, we consider
\begin{equation}
\quad \mathcal{C}=\left(\begin{array}{cc} 2 & 1 \\ 3 & 2 \end{array}\right), \label{cat2}
\end{equation}
for which the Lyapunov exponent is $\log(2+\sqrt3)$
and the stable and unstable directions are ${\bf s} = (-\sqrt3, 1)$ and ${\bf u} = ( \sqrt3, 1)$. Quantization of cat maps was first introduced in Ref. \cite{Hannay 1980} and discussed in \cite{cat}. For the case considered here, with periodic boundary conditions $\chi_q=\chi_p=0$, this results in
\begin{equation}
U^{\mathcal{C}}(Q',Q)=\sqrt{\frac{-i}{N}} e^{2i\pi(Q^{2}-Q'Q+Q'^{2})/N},
\label{pertq0}
\end{equation}
where $q=Q/N$ and $q'=Q'/N$.

We note that for the baker map the opening corresponds to a cell in the Markov partition.
As a consequence, the repeller is given in terms of an exactly self-similar fractal, the
well known middle-third Cantor set. In that respect the cat map is more generic, since
its stable and unstable manifolds intersect the opening transversally.

\section{Method and results}
\label{sec3}

Scar functions are special wavefunctions constructed by taking into account classical
information in the neighborhood of a periodic orbit
\cite{ScarFunction1,ScarFunction2,ScarFunction3,ScarFunction4,ScarFunction5,Vergini,Morpho}.
They have been developed for closed systems and are the building blocks of the
semiclassical theory of short periodic orbits, by means of which one can find eigenvalues
and eigenfunctions of a quantum system starting from purely classical quantities. For
open systems they were introduced in Ref. \cite{p1}. In this section we review their
construction and show a few examples.

%
\begin{figure}[t]
\hspace{0.0cm}
\includegraphics[angle=0.0, scale=0.4,clip]{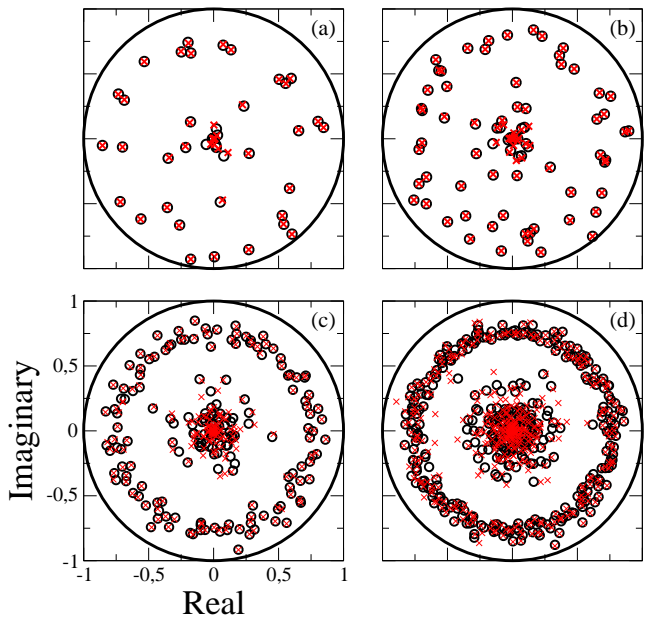}
\caption{Exact spectrum (circles) and our results (crosses), for the baker map. The pairs ($N$,$N_s$) 
consisting of the Hilbert space dimension $N$ and the number of scar functions $N_s$ are: a)($81,51$); 
b) ($177,105$); c), ($597,231$) and d) ($1821,471$). We have chosen $N_s\approx 4 N^{d/2}$ where $d$ 
is the dimension of the classical repeller. }
\label{fig3}
\end{figure}
%

%
\begin{figure}[t]
\hspace{0.0cm}
\includegraphics[angle=0.0, scale=0.5,clip]{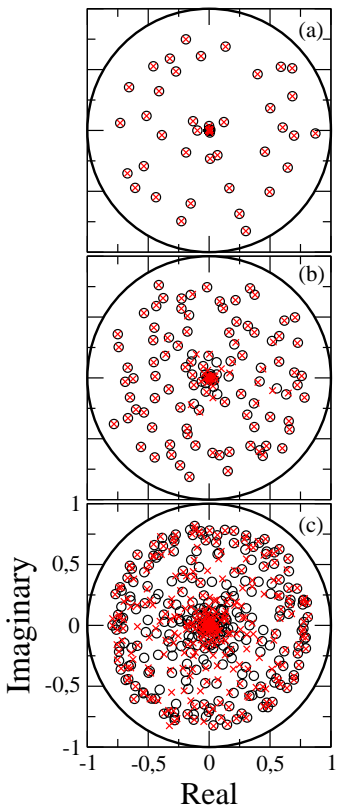}
\caption{Exact spectrum (circles) and our results (crosses), for the cat map. Here, the pairs 
($N$,$N_s$) are a)($114,90$); b) ($415,198$) and c), ($1751,476$).} 
\label{figcat}
\end{figure}
%

Let $\gamma$ be a periodic orbit of an open map (it must therefore belong to the
repeller) of fundamental period $L$, i.e. it consists of $L$ different points in the
torus: \be (q_0,p_0),(q_1,p_1),....(q_{L-1},p_{L-1}),(q_{L},p_{L})=(q_0,p_0). \ee We
associate with $\gamma$ a total of $L$ scar functions. Initially, we define coherent
states $|q_j,p_j\ra$ for each point of the orbit and a linear combination of them called
a periodic orbit mode, \be |\phi_\gamma^k\ra=\frac{1}{\sqrt{L}}\sum_{j=0}^{L-1}
\exp\{-2\pi i(j A^k_\gamma-N\theta_j)\}|q_j,p_j\ra.\ee Here $k\in\{0,\ldots,L-1\}$ and
$\theta_j=\sum_{l=0}^j S_l$, where $S_l$ is the action acquired by the $l$th coherent
state in one step of the map. The total action of the orbit is $\theta_L\equiv S_\gamma$
and $A^k_\gamma=(NS_\gamma+k)/L$.

The right and left scar functions associated with the periodic orbit are defined through
the propagation of these modes under the open map. Namely, \be\label{prop}
|\psi^R_{\gamma,k}\ra=\frac{1}{\mathcal{N}_\gamma^R}\sum_{t=0}^{\tau}
\widetilde{U}^te^{-2\pi iA^k_\gamma t}\cos\left(\frac{\pi
t}{2\tau}\right)|\phi_\gamma^k\rangle,\ee and \be
\la\psi^L_{\gamma,k}|=\frac{1}{\mathcal{N}_\gamma^L}\sum_{t=0}^{\tau}
\la\phi_\gamma^k|\widetilde{U}^te^{-2\pi iA^k_\gamma t}\cos\left(\frac{\pi
t}{2\tau}\right).\ee The constants $\mathcal{N}^{R,L}$ are chosen such that $\la
\psi_{\gamma,k}^R|\psi^R_{\gamma,k}\ra=\la \psi_{\gamma,k}^L|\psi^L_{\gamma,k}\ra$ and
$\la \psi_{\gamma,k}^L|\psi^R_{\gamma,k}\ra=1$. The cosine is used to introduce a smooth
cutoff. The time scale of the propagation, $\tau$, is taken proportional to the system's
Ehrenfest time.

%
\begin{figure}[t]
\hspace{0.0cm}
\includegraphics[angle=0.0, scale=0.35]{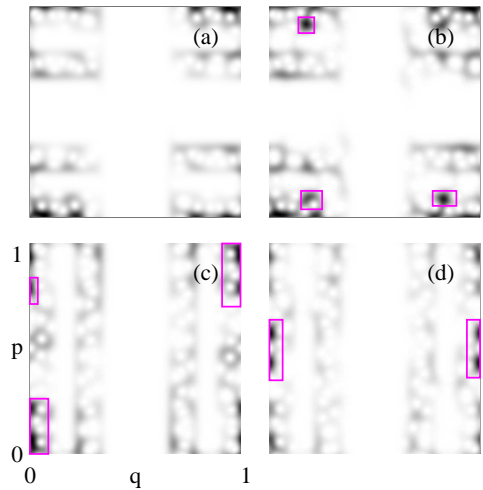}
\caption{Husimi representation of exact eigenstates and the results of our method for the baker map at 
$N=243$ and $|z|=0.895$; right eigenstate in panels a) (exact) and b) (our method), while c) and (d) 
are analogous for the left eigenstate. The overlaps between these pairs are 
$|\la \psi^{R}_{ex}|\phi^{R}_{sc}\ra|^2=0.727$ and $|\la \psi^{L}_{ex}|\phi^{L}_{sc}\ra|^2=0.556$.  
Differences are underlined by means of magenta (dark gray) rectangles.} 
\label{fig4}
\end{figure}
%

%
\begin{figure}[t]
\hspace{0.0cm}
\includegraphics[angle=0.0, scale=0.35]{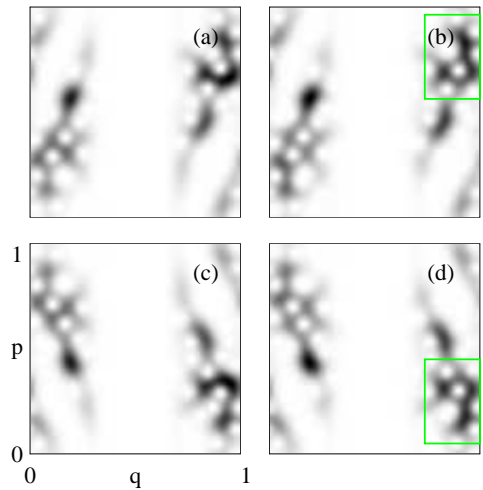}
\caption{Husimi representation of exact eigenstates and the results of our method for the cat map. Differences are 
underlined by green rectangles. 
 Right eigenstate in panels a) (exact) and b) (our method), while c) and (d) are analogous for the left eigenstate. 
Here $N=100$ and $|z|=0.722$. Overlaps are $|\la \psi^{R}_{ex}|\phi^{R}_{sc}\ra|^2=0.954$ and 
$|\la \psi^{L}_{ex}|\phi^{L}_{sc}\ra|^2=0.948$.}
\label{fhuscat}
\end{figure}
%

We use these functions to construct an approximate basis in the Hilbert space to
diagonalize our propagator. We select a number of short periodic orbits that
approximately cover the repeller, and eigenvalues and eigenfunctions are obtained by
solving a generalized eigenvalue problem \cite{p1}. In this way, we isolate the relevant
information needed to construct just the long lived resonances, without calculating the
others. According to the fractal Weyl law, the number of such resonances grows like
$N^{d/2}$ \cite{LuPrl03,SchomerusPrl04,pedrosa1,Non1,Shepelyansky08,Boro1,Hard3D}, where
$d$ is a fractal dimension of the classical repeller. Our method takes advantage of this
fact: the number of scar functions that we need to obtain a reasonable approximation to
the long-lived sector of the spectrum, denoted $N_s$, is of the order of $N^{d/2}$. The
dimension of the matrix to be diagonalized is thus substantially reduced.

Exact spectra and the results of our method are shown in Fig. \ref{fig3} for the baker
map. We notice a very clear gap developing in the spectrum, which is reproduced by the
method. This gap has been observed before \cite{Non1}, but the reasons for its existence have not yet been understood. 
The number of scar functions used was taken to scale with dimension as the fractal Weyl law: $N_s\sim N^{d/2}$, where $d$ 
is the dimension of the classical repeller. The quality of the eigenvalues obtained does not deteriorate with $N$, 
indicating that this choice is correct and the long-lived sector of the spectrum indeed has a reduced effective dimensionality.

%
\begin{figure}[t]
\hspace{0.0cm}
\includegraphics[angle=0.0, scale=0.35]{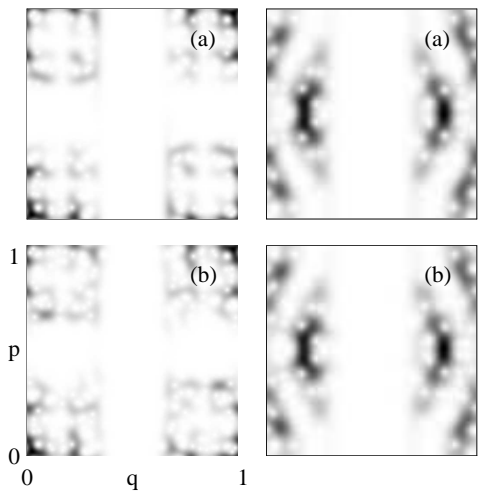}
\caption{Mixed representation $h_n(q,p)$ of the same eigenfunctions shown in Fig.
\ref{fig4} and \ref{fhuscat}, coming from: a) exact calculation; b) our method. 
Results for the baker map are on the left and for the cat map are on the right. The agreement is much
better than in Fig. \ref{fig4} and \ref{fhuscat}, for both systems.} 
\label{fig5}
\end{figure}
%

Analogous results are presented for the cat map in Fig. \ref{figcat}. The quality of
the individual eigenvalues is not as good as in the baker map. However, the scaling is
preserved, i.e. the results confirm that we only need about $N^{d/2}$ scar functions for
each value of $N$ in order to reproduce the portion of the spectrum lying closest to the
unit circle.

It has been shown \cite{KeatingPrl06} that the right eigenstates are located in the
unstable manifold of the repeller, in the sense that their Husimi representations
$\Psi_n^R\mapsto|\la q,p|\Psi_n^R\ra|^2$ are nearly zero outside that set. On the other
hand, left eigenstates are supported by the stable manifold. Our method is based on
functions that are approximately supported by the repeller. Therefore, perhaps
surprisingly, although it is able to provide very accurate eigenvalues, the corresponding eigenfunctions are not necessarily well reproduced. 
We show in Fig. \ref{fig4} the right and
left eigenfunctions of the eigenvalue $z=0.895$ of the baker map for $N=243$. 
We see that the results from our method have significant differences compared to the exact ones, in
regions away from the repeller. We also show analogous results for the cat
map ($z=0.722$, $N=100$) in Fig. \ref{fhuscat}. Differences are not as noticeable in this case.

The method is, however, able to accurately reproduce the mixed representation
$h_n(q,p)=|\la q,p|\Psi_n^R\ra\la \Psi_n^L|q,p\ra|^2$ which was recently introduced in
\cite{ermannPrl09}. The reason for that is that this phase space quantity is supported on the 
intersection of the individual supports, which is precisely the repeller where our
scar functions live. In Fig. \ref{fig5} we can clearly see that the values of $h_n(q,p)$
computed with our method indeed coincide with the exact ones for both systems.

\section{Conclusions}
\label{sec5} We have shown in two paradigmatic models that our recently developed method
\cite{p1} to obtain quantum resonances of chaotic systems from their classical properties indeed shows the same scaling as the fractal Weyl law. Namely, only a fraction $N^{d/2}$
(where $d$ is the dimension of the classical repeller) of our special basis states are
required to provide a good approximation to the long-lived sector of the spectrum. When
the Hilbert space dimension $N$ becomes large, this actually represents only a small
fraction of $N$. We found that the method gives good but not excellent approximations to
the right and left resonant eigenstates independently, because of the very
limited support of the basis states we use. However, a mixed quantity involving both right and left eigenstates can be successfully obtained. Perhaps with some modification the method could also reproduce them, but further investigation is needed.

\section*{Acknowledgments}

Fruitful discussions with Marcos Saraceno and Eduardo Vergini are gratefully acknowledged.



\end{document}